\documentclass[a4paper,preprint,12pt]{elsarticle}
\usepackage{amsmath,amsthm,amssymb}
\usepackage{graphicx}
\usepackage{lineno} 
\usepackage{multirow}
\usepackage[margin=0.8in]{geometry}
\usepackage[colorlinks]{hyperref}

\journal{arXiv}  

\begin{document}

\begin{frontmatter}

\title{\textit{In situ} analysis of catalyst composition during gold catalyzed GaAs nanowire growth}

\author[add1,add2]{Carina B. Maliakkal$^*$}
\ead{carina\_babu.maliakkal$@$ftf.lth.se} 
\author[add1,add3]{Daniel Jacobsson}
\author[add1,add2]{Marcus Tornberg}
\author[add1,add3]{Axel R. Persson}
\author[add1,add2]{Jonas Johansson}
\author[add1,add3]{Reine Wallenberg}
\author[add1,add2,add3]{Kimberly A. Dick}

\address[add1]{NanoLund, Lund University, 22100, Lund, Sweden.} 
\address[add2]{Solid State Physics, Lund University, Box 118, 22100, Lund, Sweden.}
\address[add3]{National Center for High Resolution Electron Microscopy and Centre for Analysis and Synthesis, Lund University, Box 124, 22100, Lund, Sweden.}

\begin{abstract}  
Semiconductor nanowires offer the opportunity to incorporate novel structures and functionality into electronic and optoelectronic devices. A clear understanding of the nanowire growth mechanism is essential for well-controlled growth of structures with desired properties, but the understanding is currently limited by a lack of empirical measurements of important parameters during growth, such as catalyst particle composition. However, this is difficult to accurately determine by investigating post-growth. 
We report direct measurement of the catalyst composition of individual gold seeded GaAs nanowires inside an electron microscope as they grow. 
The Ga content in the catalyst during growth increased with both temperature and Ga precursor flux. A direct comparison of the calculated phase diagrams of the Au-Ga-As ternary system to the measured catalyst composition not only lets us estimate the As content in the catalyst but also indicates the relevance of phase diagrams to understanding nanowire growth.    
\end{abstract}


\end{frontmatter}

\section*{Introduction}

Nanowire growth by the vapor-liquid-solid (VLS) method is an important technique for producing well-controlled nanocrystals suitable for quantum components. For III-V semiconductors, an important materials system for future technologies within electronics, solid-state lighting and quantum processing, VLS growth enables the fabrication of, for example, lattice-mismatched heterostructures,\cite{guo2006structural, larsson2006strain, caroff2009insb} metastable crystal phases and crystal phase tuning,\cite{persson_solid-phase_2004, jacobs2007electronic, joyce_phase_2010, dick_control_2010} and unusual ternary alloys\cite{sukrittanon2014growth,namazi2017direct}. VLS growth has been well-studied for over two decades, and extensive theoretical efforts exist to explain the growth process itself,\cite{dubrovskii_refinement_2017, schwarz_elementary_2011, nebolsin_role_2003, krogstrup_advances_2013, glas_chemical_2010} the observed trends with experimental parameters,\cite{dubrovskii_group_2016, plante_analytical_2009, ramdani_arsenic_2013, dubrovskii_diffusion-controlled_2007} and the existence of metastable structures\cite{glas_why_2007, dubrovskii_growth_2008, johansson2015polytype}. 
However, validation of theoretical predictions remains extremely difficult due to the large number of variable material properties and accessible experimental parameters, and the subsequent variance in reported experimental trends. Additionally, many of the important fundamental parameters, such as surface and interface energies for relevant growth conditions, are unknown.\cite{tornberg2017thermodynamic,panciera_controlling_2016} 
Consequently, there are a wide range of competing and complementary models that can explain observations such as crystal phase trends\cite{joyce_phase_2010,dubrovskii_mono-_2015,krogstrup_impact_2011,algra_formation_2011} and diameter-growth rate dependencies\cite{dubrovskii_general_2007,froberg2007diameter,li2013nucleation, borgstrom2007synergetic, li2013readsorption, borg2013geometric}. 

The use of \textit{in situ} characterization methods to gain direct insights into semiconductor nanowire growth in real time is one of the most effective ways to refine theoretical explanations and predictions, and in turn to better understand the conditions needed to design these materials with high control.\cite{ek_atomic-scale_2018} Examples of previous \textit{in situ} studies include X-ray diffraction to understand phase and structural evolution,\cite{ kirkham_tracking_2010, schroth_evolution_2015, krogstrup_situ_2012} infrared spectroscopy to correlate surface chemistry during growth with resulting nanowire morphology,\cite{sivaram_surface_2016, sivaram2015direct, shin2014interplay} reflectance high-energy electron diffraction to follow nucleation and structural changes,\cite{tchernycheva2006temperature, xu2012faceting, jo2018real} optical reflectometry to monitor growth rate evolution,\cite{ heurlin_situ_2015, clement2006situ} and mass spectrometry to study nucleation\cite{fernandez_monitoring_2015}. In addition, \textit{in situ} scanning electron microscopy has been used to directly follow nanowire growth and morphology, and combined with Auger electron spectroscopy to track surface chemistry.\cite{kolibal_synergic_2016} Finally, \textit{in situ} transmission electron microscopy (TEM) has proven to be one of the most powerful ways to gain insight into nanowire growth in a directly interpretable way. Importantly, the information provided by this method is local, meaning that the nanowire, the growth-enabling liquid droplet and the interface between them can be visually identified and independently studied. This method has led to numerous significant breakthroughs in nanowire growth such as vapor-solid-solid growth,\cite{kodambaka_germanium_2007,hofmann_ledge-flow-controlled_2008} corner truncation,\cite{wen_periodically_2011,gamalski2011cyclic} and step-flow\cite{hofmann_ledge-flow-controlled_2008}.
\textit{In situ} TEM studies have been particularly beneficial for understanding 
bilayer growth kinetics,\cite{chou_atomic-scale_2014}
crystal phase switching,\cite{jacobsson_interface_2016} 
triple-phase-line nucleation,\cite{harmand_atomic_2018}	
and double bilayer growth\cite{gamalski_atomic_2016} in III-V nanowires.

One essential thing that remains to be investigated is the local composition of the nanostructure during growth. This is necessary to understand composition evolution in for instance heterostructure and ternary nanowires, but even more importantly, it is necessary for understanding the composition of the liquid metal droplet as a function of growth parameters and how this is correlated with the resulting nanowire properties.\cite{dubrovskii_composition_2018} The composition of the catalyst particle is a pivotal factor that determines its thermodynamic parameters such as vapor pressure, chemical potential and surface energies which in turn decide the nanowire structure, growth rate, composition etc.\cite{dubrovskii_zeldovich_2015, dubrovskii_influence_2014, leshchenko_nucleation-limited_2018}. 
So far, the composition of the catalyst particles have been measured post-growth and was shown to depend on the conditions used for terminating the growth\cite{harmand_analysis_2005, jacobsson_zincblende--wurtzite_2013} (more details in Supplementary Information section S1), implying that post-growth composition of the particle is different from its composition during growth.
To our knowledge, there has been no direct \textit{in situ} measurement of catalyst composition during nanowire growth. An indirect estimation of the Au-Ga catalyst composition during GaAs nanowire growth has been reported by comparing the dimensions of the starting Au seeds particles and the catalyst during growth by assuming that the seed material does not diffuse out of the catalyst particle.\cite{jacobsson_interface_2016} By comparing the catalyst composition measured \textit{in situ} with calculated phase diagram one can also investigate the validity of using equilibrium phase diagrams in understanding the non-equilibrium nanowire growth. In nanowire growth models which consider phase diagrams, the nanowire growth is assumed to happen when the composition exceeds the liquidus line along which the nanowire material coexists in equilibrium with a stable liquid catalyst alloy.\cite{chatillon_thermodynamics_2009} 

In this article we report for the first time direct \textit{in situ} measurement of catalyst composition during nanowire growth. 
We use \textit{in situ} X-ray energy dispersive spectroscopy (XEDS) combined with \textit{in situ} TEM to investigate the composition of the metal droplet as a function of growth parameters for Au-seeded GaAs nanowires grown by the VLS method. We show that the droplet consists of a significant quantity of Ga for all growth conditions, which increases with temperature for constant precursor flow. We do not observe any As significantly above the detection limit of the XEDS technique. Using calculated ternary phase diagrams, we show that the As content can however be estimated based on the temperature dependence of the Ga content. We also observe that the Ga content of the droplet for a given temperature is relatively independent of the ratio of V/III precursor species, so long as this ratio is above a certain threshold. Below this threshold, the Ga content increases strongly with decreasing V/III ratio, accompanied by a volume increase in the droplet. We show that the droplet volume scales with the Ga content, validating earlier works that used volume as an estimate of Ga.\cite{jacobsson_interface_2016} The trend with V/III ratio is understood to correlate with a gradual transition towards the so-called ´´arsenic-limited´´ growth, whereby the droplet initially swells up but eventually stabilizes.\cite{tersoff_stable_2015} The results demonstrate that \textit{in situ} XEDS is a useful and direct way to gain important insight and information on nanowire growth in real time, and will be similarly appropriate for other types of processes occurring at similar temperatures and overall gas pressures. 
Finally, the measurements of the droplet composition as a function of growth parameters will provide important inputs to validation and modification of theoretical models describing nanowire growth.\cite{glas_chemical_2010,dubrovskii_composition_2018,tizei_iiiv_2009}

\section*{Results and discussion}

\subsection*{Growth of GaAs nanowires and \textit{in situ} measurements}
Au nanoparticles deposited on a silicon nitride-based heating grid were used to seed the nanowire growth. Nanowires were grown inside a Hitachi HF3300S environmental TEM integrated with a custom metal organic chemical vapor deposition (MOCVD) system. Trimethylgallium (TMGa) and arsine (AsH$_3$), which are the most common precursor gases in MOCVD growth of GaAs, were used for this study. The chemical composition of the catalyst was studied by XEDS as a function of temperature and the ratio of precursor fluxes, which are two very important parameters in typical MOCVD-growth. Please refer to Methods section for more details.

\subsection*{XEDS of catalyst measured \textit{in situ}}

\begin{figure}[!h]
\centerline {\includegraphics[width=88mm]{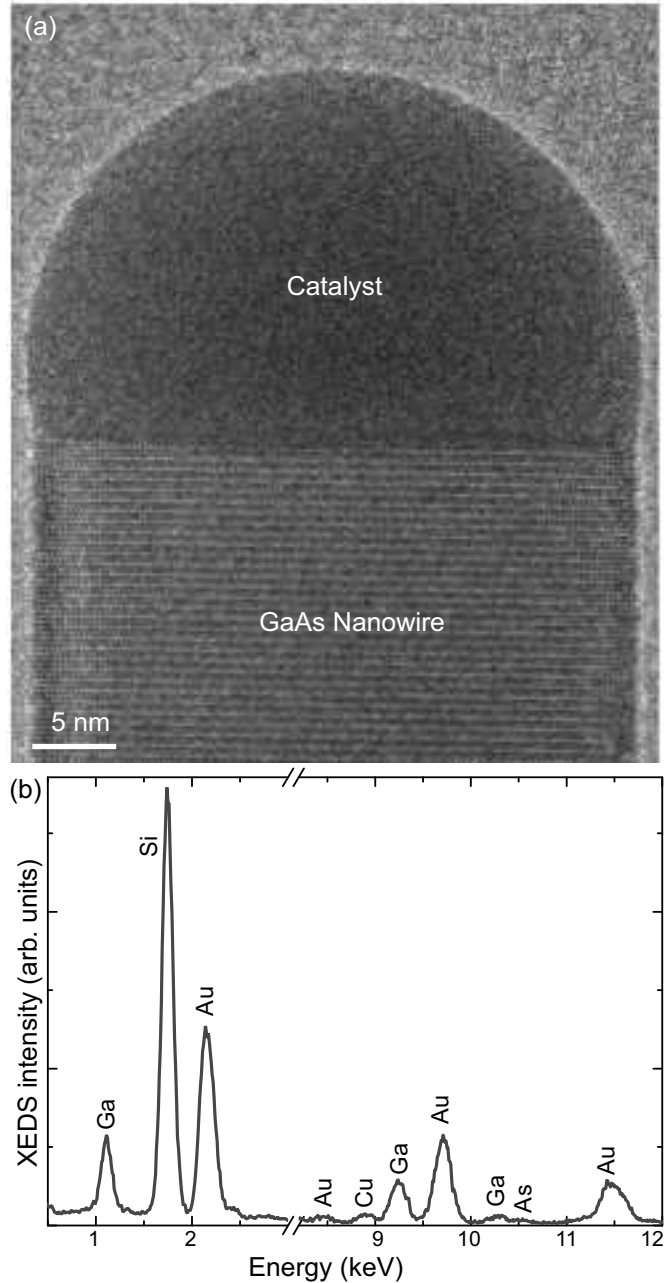}}
\caption{\label{fig:edxeg}\textbf{\textit{In situ} Au-seeded GaAs nanowire growth:} (a) TEM image of a GaAs nanowire growing inside the TEM on a SiN$_x$ grid at 440~$^\circ$C and V/III ratio of 1300. (b) XEDS spectrum of the catalyst particle at the same conditions measured \textit{in situ} during its growth. The atomic species giving rise to the different peaks is indicated in the plot. }
\end{figure}

The catalyst composition was measured using XEDS \textit{in situ} as the GaAs nanowire grows. An example XEDS spectrum is shown in Fig. \ref{fig:edxeg} (b) from the nanowire shown in Fig. \ref{fig:edxeg} (a). Clear signals from Ga and Au in the catalyst are observed for all spectra, along with a strong Si signal arising from the SiN$_x$ grid on which the nanowires are growing, and system artefacts such as Cu arising from microscope components. Quantification of the XEDS spectrum in Fig. \ref{fig:edxeg} (b), measured at 440~$^\circ$C, gives a Ga:Au atomic ratio of about 30:70 (assuming only Ga and Au are present). Some spectra also show small features that could be due to As, but the quantity is too low to be conclusively attributed to As and quantified (further information is available in the Methods section). Since any As within the catalyst is too low to be directly quantified, we can only put an upper limit on the As content (of approximately 1-2 atomic \%). Our observation of very low concentration of As in the catalyst during growth is consistent with theoretical calculations; for instance, Glas \textit{et al.} calculate the As content to be about 1~\% (depending on radius and contact angle).\cite{glas2013predictive}  M\aa rtensson \textit{et al.} predicted As content to be roughly in the range 0.01~\% - 1~\% depending on the growth conditions.\cite{maartensson2019simulation} Post-growth XEDS reports have also claimed As to be below detection limits.\cite{jacobsson_zincblende--wurtzite_2013}

\vspace{1.5cm}
\subsection*{Catalyst composition as a function of temperature}

\begin{figure}
\centerline {\includegraphics[width=88mm]{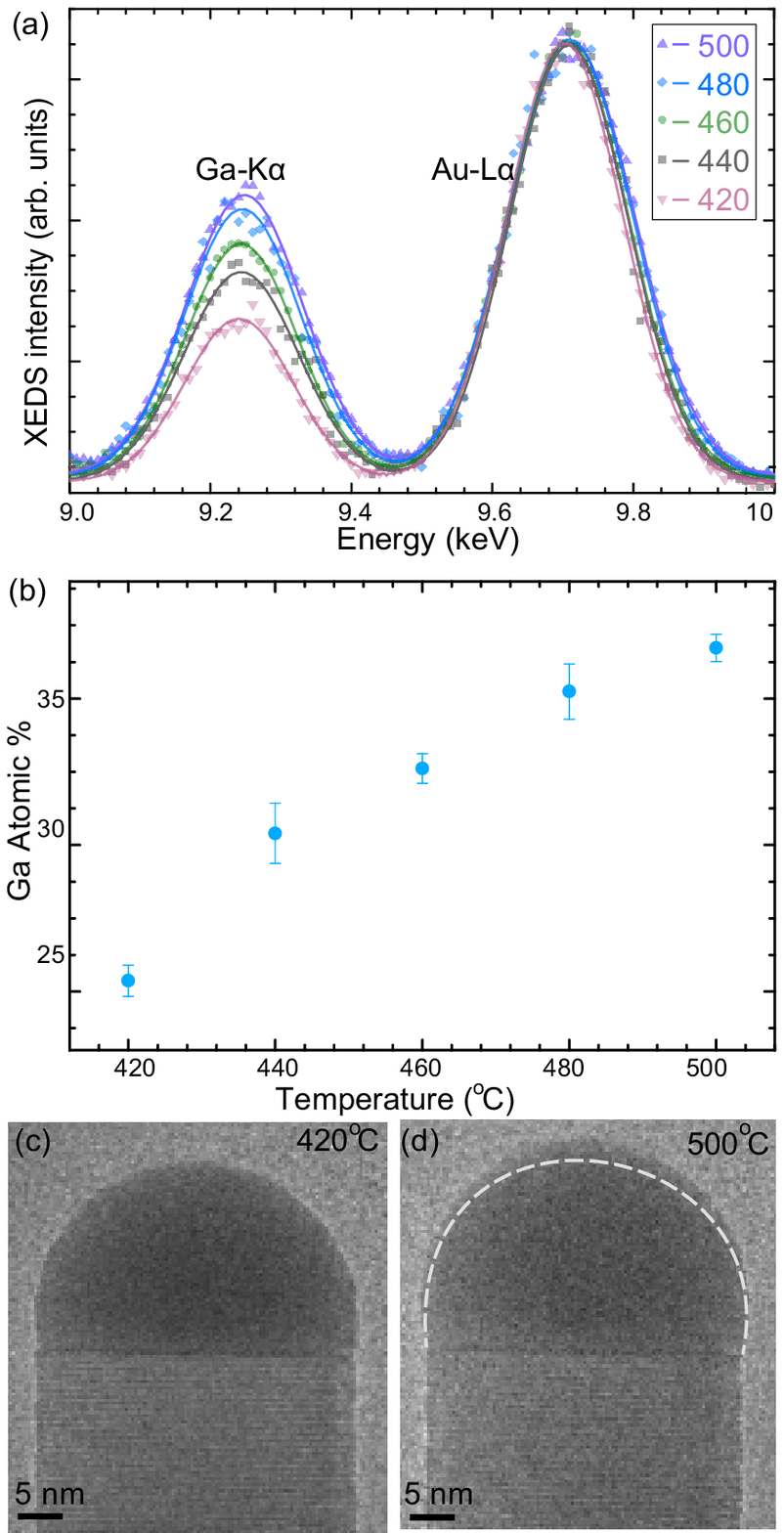}}
\caption{\label{fig:temp}\textbf{Catalyst composition as a function of temperature:} (a) The Ga-K$\alpha$ and Au-L$\alpha$ peaks in XEDS spectra measured at different temperatures (in $^\circ$C). The spectra are normalized with respect to the Au-L$\alpha$ peak. The Ga peak intensity increases relative to the Au peak with increasing temperature. (b) The atomic percentage of Ga in the catalyst particle measured as a function of temperature at constant V/III ratio. With increasing temperature, the catalyst stabilizes with more Ga resulting in larger catalyst particle as seen in (c) and (d). TEM images of the catalyst at 420~$^\circ$C (c) and 500~$^\circ$C (d). At 500~$^\circ$C the catalyst is larger than at 420~$^\circ$C. The outline of the catalyst at 420~$^\circ$C is depicted on top of the TEM image at 500~$^\circ$C (d) by a white dashed line. }
\end{figure}

The composition of the catalyst particle during nanowire growth was measured as a function of temperature and is shown in Fig. \ref{fig:temp} . During this experiment the Ga precursor flow was set to be relatively low (V/III = 1300; the relevance of this choice will be discussed later). More details of the experiment can be found in the Methods. The Ga content in the catalyst increases with temperature as can be observed in the XEDS spectra normalized to the Au-L$\alpha$ peak (Fig. \ref{fig:temp} (a)). Quantification of the XEDS spectra shows that when the temperature was increased monotonously from 440~$^\circ$C to 500~$^\circ$C the Ga content increased from 30 to 36 atomic~\%  (Fig. \ref{fig:temp} (b)). The temperature was then decreased to 420~$^\circ$C, after which the Ga content decreased. A small but measurable change in the volume of the catalyst droplet at different temperatures was observed, as illustrated in Fig. \ref{fig:temp} (c), indicating that the composition change is due primarily to an increase in Ga rather than a loss of Au atoms. (The correlation between Ga concentration and catalyst volume is discussed in more detail in Supplementary Information section S6)

\begin{figure}
\centerline {\includegraphics[width=88mm]{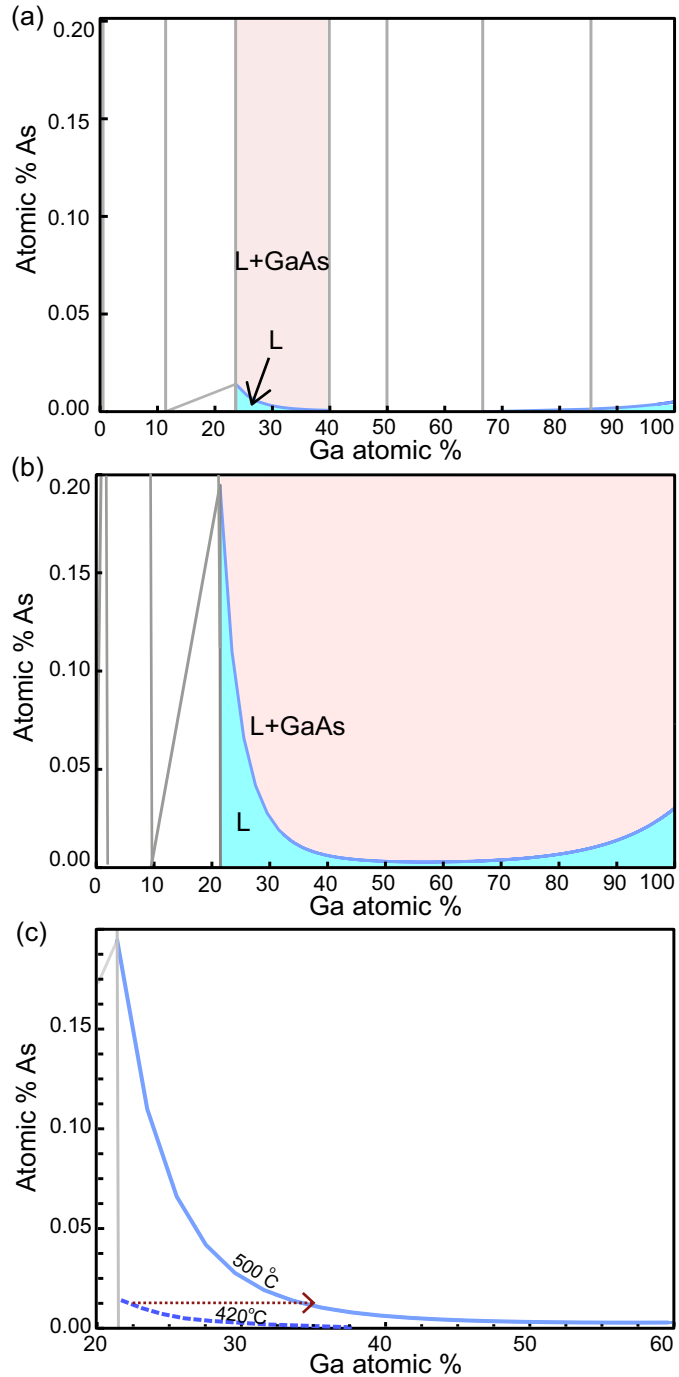}}
\caption{\label{fig:PDT}\textbf{Au-Ga-As ternary phase diagrams:} (a) Au-Ga-As phase diagram calculated at a fixed temperature of 420~$^\circ$C. The liquidus is shown as a blue curve. The region with a stable Au-Ga-As liquid is indicated as `L'.
(b) Au-Ga-As phase diagram calculated for 500~$^\circ$C (c) Magnified section of phase diagram at 500~$^\circ$C overlayed with the liquidus line at 420~$^\circ$C is indicated by the blue dashed curve. A representative constant As condition is indicated as a horizontal brown dotted line, showing that the Ga concentration at the liquidus line increases with increasing temperature.
}
\end{figure}

To understand the increase of Ga content in the catalyst with temperature we now consider some representative phase diagrams of the Au-Ga-As ternary alloy. Although nanowire growth is of course not an equilibrium process, phase diagrams can still be very useful in understanding the process by giving a visualization of the thermodynamic reference state of the system.\cite{ghasemi_phase_2017} Fig. \ref{fig:PDT} (a) and (b) shows the ternary Au-Ga-As isothermal phase diagrams for two temperatures- 420~$^\circ$C and 500~$^\circ$C, calculated with the Thermo-Calc software using the thermodynamic data assessed by Ghasemi \textit{et al.}\cite{ghasemi_phase_2017} Note that the range of the plots shown here reaches 100~\% for Ga and Au (x-axis), but only 0.2~\% for As (y-axis). This range was chosen to clearly see the regions relevant for nanowire growth. 
Since for As $>$ 0.2~\% there are no new phases formed compared to the ones already seen in Fig. \ref{fig:PDT} (a), (b), the phase diagram can be just extrapolated linearly for that regime. 
According to the phase diagram in Fig. \ref{fig:PDT} (a), a single phase stable liquid alloy of Au-Ga-As exists in a very narrow range of the ternary composition near As 0.01~\%, Ga 26~\% and Au 74~\%. At 500~$^\circ$C this region is broader. For the nanowire material to nucleate or precipitate out of the liquid catalyst alloy, the liquid catalyst alloy should first cross the liquidus line and reach a supersaturated/supercooled state.\cite{ghasemi_phase_2017} Supersaturation in this scenario means that there would be more Ga and As in the catalyst than in a corresponding alloy at thermodynamic equilibrium. 
 The maximum allowed As or Ga content, such that the alloy remains a liquid at equilibrium, is given by the liquidus line (blue line in Fig. \ref{fig:PDT}). In other words, the liquidus separates the stable liquid alloy and a two-phase region with coexistence of solid GaAs and a Au-Ga-As ternary liquid. A supersaturated liquid alloy would hence be slightly above or on the right side of the blue liquidus line in Fig. \ref{fig:PDT} (keeping in mind that the number of excess atoms required to form an entire GaAs bilayer would correspond to about 1~\% in the Au droplet for the catalyst/nanowire dimensions discussed here). So for a basic understanding of the process let us consider that the nanowire growth happens at (or more precisely, immediately beyond) the liquidus line.

The liquidus shifts to higher Ga content at higher temperature (Fig. \ref{fig:PDT}). Fig. \ref{fig:PDT} (c) shows a magnified section of Fig. \ref{fig:PDT} (b) (at 500~$^\circ$C) where the liquidus at 420~$^\circ$C is overlayed on it as a blue dashed curve. 
In principle, when the temperature is changed both the Ga and As concentrations in the catalyst could change. However, the As concentration in the catalyst would equilibrate with the ambient AsH$_3$ partial pressure.\cite{maartensson2019simulation} Since in this experiment the AsH$_3$ flow is fixed, the As content in the catalyst would be mostly invariant with temperature.

\begin{figure}
\centerline {\includegraphics[width=88mm]{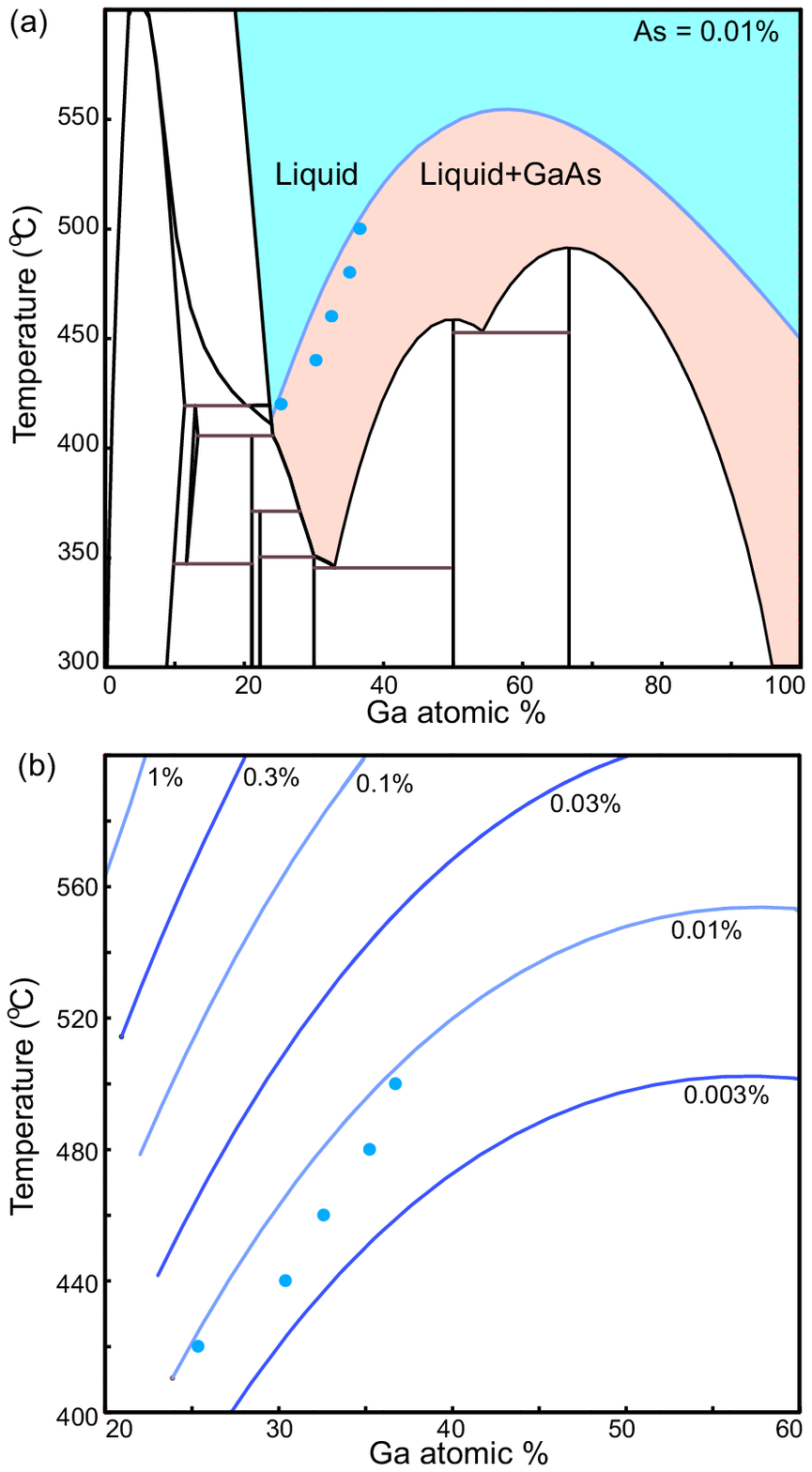}}
\caption{\label{fig:PDAs}\textbf{Comparison of measured Ga with constant As phase diagram sections}: (a) Au$_{99.99\%}$As$_{0.01\%}$- Ga$_{99.99\%}$As$_{0.01\%}$ phase diagram (or projection of the Au-Ga-As phase diagram at a fixed As content of 0.01~\%). The choice of this concentration of As was such that the measured Ga percentage lies in the supersaturation regime (i.e. right side or below the liquidus line shown by the blue curve.) Blue dots are the experimentally measured Ga content. (b) Liquidus line calculated for different As concentrations. The As concentration is labelled on each liquidus.}
\end{figure}

Let us now look at calculated phase diagrams at a few different As concentrations.
We see that the Ga concentration measured experimentally by XEDS at different temperatures (blue dots in Fig. \ref{fig:PDAs} (a)) is in close agreement with Ga concentration of the liquidus line calculated for As = 0.01~\%. This close agreement in liquidus shape and experimental data indicates that the As content likely does not change very much with temperature, since this would result in a steeper temperature dependence of the liquidus (less change of Ga with temperature) according to Fig. \ref{fig:PDT} (c). The requirement of supersaturation implies that the measured composition should be such that Ga is more than the corresponding equilibrium concentration and hence on the right side of the liquidus line (blue). It is clear that As~\% cannot be much lower than 0.01~\% while still fulfilling this condition: for instance As $=$ 0.003~\% would imply that the Ga concentration is not sufficient to supersaturate the catalyst particle, as the measured Ga content would be less than the equilibrium Ga content (the liquidus line concentration) (Fig. \ref{fig:PDAs} (b)). 

On the other hand, for higher As, the liquidus line moves upwards in temperature, which limits how high the As content can be in a fully liquid catalyst at equilibrium. 
The Ga concentration varies during growth and is highest at nucleation and lowest after completion of a layer. When the Ga is at its lowest, it is likely that it reaches equilibrium. If this equilibrium does not correspond to a liquidus, there is a high chance that solid gold-containing phases form, and this we do not observe. An upper limit of the As concentration of 0.3 \% can be estimated such that a liquidus exists at the investigated temperatures.
The full phase diagram calculated for As of 0.3~\% can be found in the Supplementary Information section S5. Note that in the experiment discussed, the temperature was varied from 440~$^\circ$C to 500~$^\circ$C by increasing the temperature in steps, so no super-cooling effects would appear in this range. Additionally, we must emphasize that since the XEDS spectra were collected for several minutes, the measured composition represents the average over time rather than the maximum supersaturation reached (which would correspond to the supersaturation at the nucleation of a new layer). 

 	In short we estimate the average As concentration in this Au-Ga-As liquid alloy to be in the order of 0.01\% (or in the range 0.01~\%$<$As$<$0.3~\%) in the temperature ranges typically used for GaAs nanowire growth in the high V/III regime. Although this value is low, it is consistent with theoretical estimates for such low AsH$_3$ flows as used here.\cite{maartensson2019simulation} Moreover, 0.01~\% As in a droplet of this size corresponds to around 100 atoms, which is sufficient to form a critical nucleus to initiate a new GaAs bilayer. We note that this As content may also increase slightly with temperature, but the change should still fall within this overall range. This study also illustrates the value of phase diagrams to aid in the qualitative understanding of nanowire growth mechanisms.

\subsection*{Catalyst composition as a function of precursor flux}
	
\begin{figure}
\centerline {\includegraphics[width=88mm]{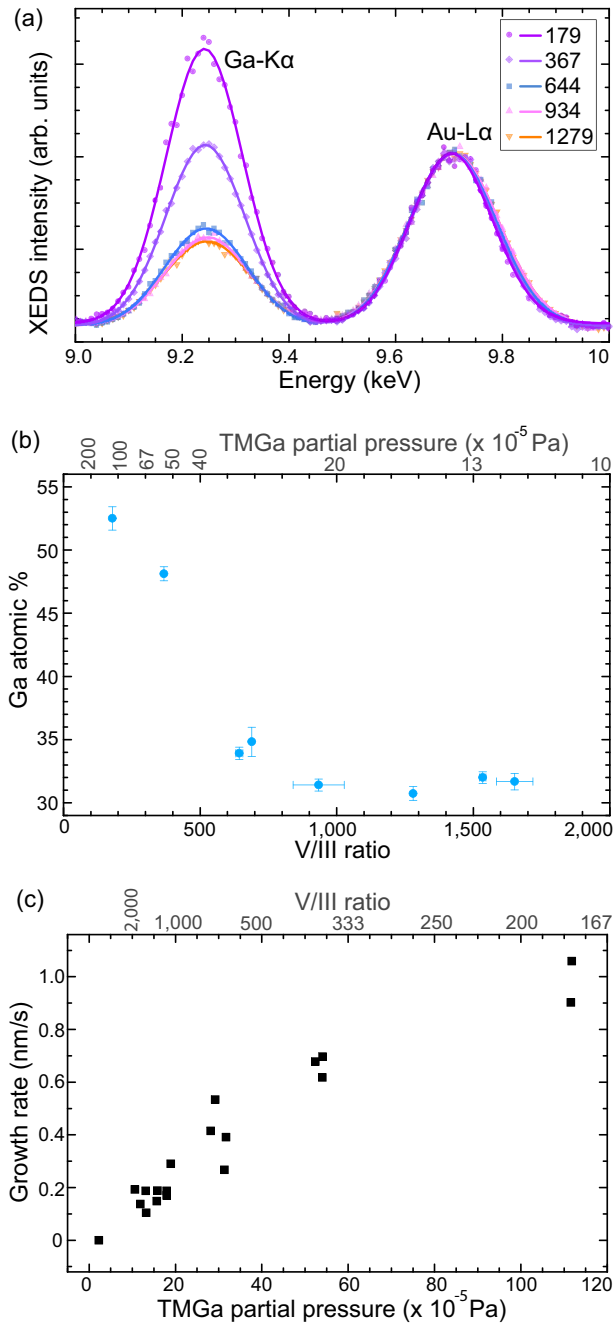}}
\caption{\label{fig:VIIIedx}\textbf{Catalyst composition and growth rate as a function of V/III ratio:} (a) The Ga-K$\alpha$ and Au-L$\alpha$ peaks in XEDS spectra measured at different V/III ratios. The signal is normalized with respect to the Au-L$\alpha$ peak. The Ga peak intensity relative to the Au peak increases with decreasing V/III ratio. (b) The Ga\% in the catalyst particle measured for varying V/III ratio. (c) Average growth rate measured from the videos plotted as a function of TMGa partial pressure. The background gradient color in (b) and (c) is such that purple indicates high TMGa regime while orange indicates low TMGa regime. The top axis in (b) and (c) is nonlinear.
}
\end{figure}	

V/III ratio i.e. the ratio of the group V precursor flux (AsH$_3$ here) to the group III precursor (TMGa here), is a very important parameter for the growth of III-V semiconductors. We now discuss the change of catalyst composition as a function of V/III ratio, at a fixed temperature of 420~$^\circ$C measured on another nanowire (Fig. \ref{fig:VIIIedx} (a), (b)). (Please see Methods section for details on how the V/III ratio is measured in this experiment.) We had set the AsH$_3$ flow to be fixed and changed only the TMGa flow in this experiment. When the TMGa flow was stopped the nanowire was neither growing nor etching; the Ga content in the catalyst was measured to be 27~\% by XEDS. (At elevated temperatures GaAs gets slowly etched by the Au-based catalyst if the precursors are not supplied appropriately.\cite{persson_solid-phase_2004}) 
 The Ga concentration in the catalyst then increases monotonically with increasing Ga precursor flux. At high V/III ratios (or low TMGa) a small increase of TMGa does not change the Ga concentration much; in fact, the Ga concentration is effectively constant within the resolution limit of the XEDS measurement. Since the Ga content is in a steady state (between incoming Ga, controlled by the TMGa flow, and outgoing Ga, primarily controlled by the nanowire growth), this result suggests that the nanowire growth is mainly limited by the TMGa flow in this regime. This is the regime that we used for studying the effect of temperature on the catalyst composition, and serves as further confirmation that a thermodynamic (rather than kinetic) understanding of the temperature trend is appropriate.

	At low V/III ratios i.e. below about 900 (Fig. \ref{fig:VIIIedx} (b)), the Ga concentration increases rapidly with increasing TMGa. (A similar increase of Ga content at lower V/III ratio was seen at 500~$^\circ$C also as is shown in Supplementary Information section S4.) 
This trend in the measured composition by XEDS is accompanied by a large, clearly visible increase in the size of the catalyst droplet. Similar swelling of the catalyst at low V/III has previously been observed both \textit{ex situ}\cite{lehmann2015crystal,dheeraj2012controlling} and \textit{in situ}\cite{chou_atomic-scale_2014,jacobsson_interface_2016}, and has also been predicted theoretically\cite{dubrovskii2018composition}. Previous reports have associated this effect with a transition to a ‘V-limited regime’ where the Ga supply is effectively higher than the As supply, and so excess Ga accumulates in the droplet before reaching a new quasi-steady-state composition.\cite{chou_atomic-scale_2014,jacobsson_interface_2016} If we consider this in terms of the phase diagram in Fig. \ref{fig:PDT} (a), the results indicate that the As concentration of the droplet is constant when TMGa is low (equilibrating with the As in the vapor as predicted theoretically\cite{maartensson2019simulation}). For high TMGa, the increased Ga in the droplet may suggest that the As concentration decreases with decreasing V/III ratio.

\begin{figure}
\centerline {\includegraphics[width=88mm]{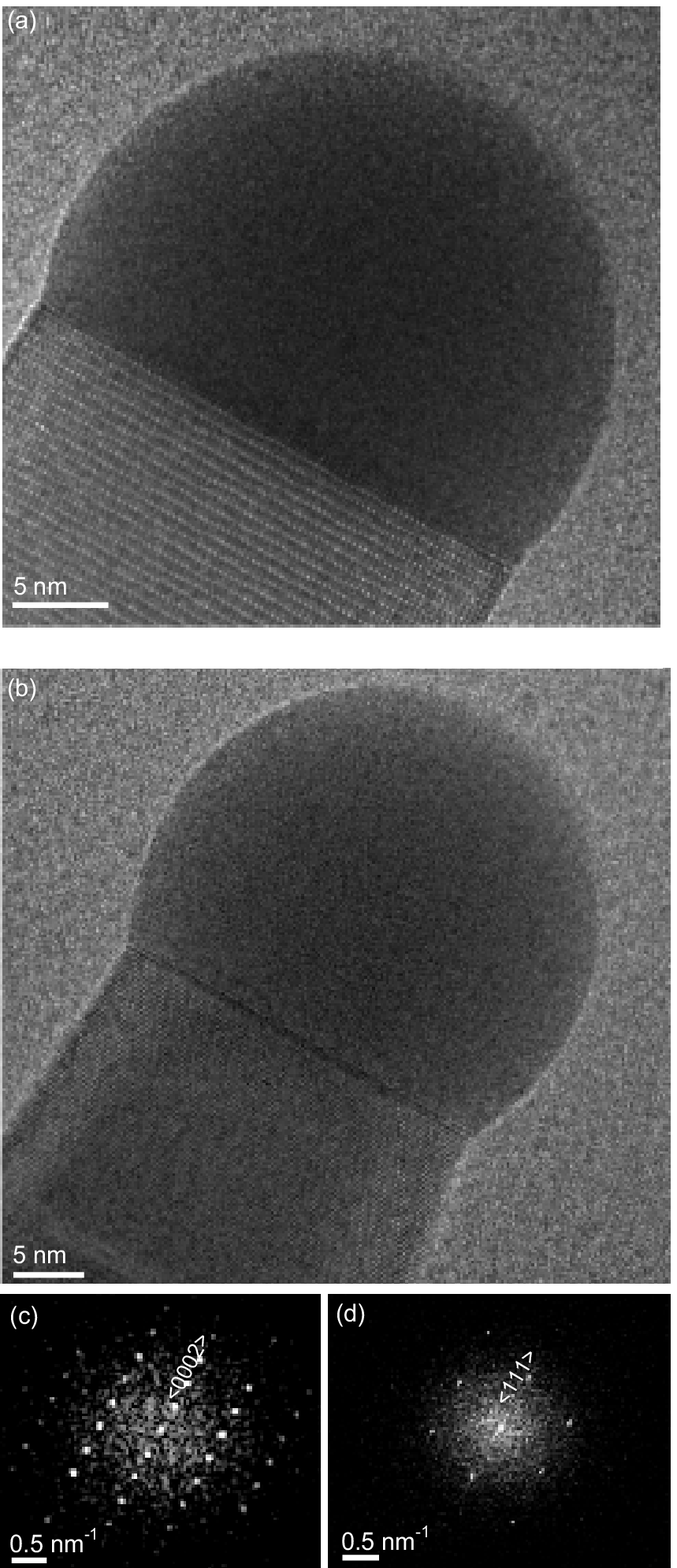}}
\caption{\label{fig:ViiiTEMs}\textbf{Catalyst morphology and nanowire structure:} (a) TEM image at high V/III ratios where the nanowire is growing in the wurtzite structure. (b) TEM image at low V/III ratios where the nanowire is growing in the zincblende structure. (c) and (d) are processed reduced FFTs of (a) and (b) respectively, showing wurtzite and zincblende structure respectively. 
}
\end{figure}

	The average growth rate of the nanowire as a function of TMGa partial pressure (and V/III ratio) is shown in Fig. \ref{fig:VIIIedx} (c). It is clear that the growth rate increases with TMGa flow over the full range. The trend is not linear however: for low TMGa (high V/III), there is a steep increase, but at higher TMGa (low V/III) this trend slows, potentially saturating at very high TMGa. The apparently linear trend between growth rate and TMGa for low TMGa flow is consistent with our interpretation above that the (effectively) constant Ga concentration in the catalyst is a consequence of Ga limiting the growth rate. Following the reasoning of M\aa rtensson \textit{et al.}\cite{maartensson2019simulation}, we conclude that when AsH$_3$ is very much in excess, the As concentration in the catalyst quickly reaches a maximum concentration which is in steady state with re-evaporation to the vapor; so long as the nucleation barrier does not shift significantly with growth parameters, the growth rate is limited by the time required for the Ga concentration to reach the level needed to overcome the nucleation barrier. The weakening of this trend at high TMGa, where increased Ga is observed in the catalyst, indicates a transition to a growth regime where As plays a limiting role. In this regime, the high TMGa flow allows the Ga concentration to exceed the value reached in the high V/III regime, before the As reaches the concentration that would be in steady state with vapor. Since the supersaturation is determined by both Ga and As species, the nucleation barrier is then reached for higher Ga and lower As concentrations (determined by the V/III ratio). The growth rate will then depend on both TMGa and AsH$_3$ flows. True As-limited growth would be predicted for even higher TMGa flows, although it is not clear whether such a regime could actually be reached in experiments. 

	In addition to the change of the droplet size with V/III ratio, a change in the crystal structure of the nanowire was also observed. In the high V/III ratio regime (about 800-1700) the nanowire grew in the wurtzite structure along the $<$0001$>$ direction (Fig. \ref{fig:ViiiTEMs} (a), (c)). When the V/III ratio was decreased, stacking faults started to appear in the wurtzite nanowire. At even lower V/III ratios close to 350 the nanowire grew as a mixture of both zincblende and wurtzite structures. At still lower V/III ratio of about 180, the same nanowire grew in the zincblende structure along the $<$111$>$ axis (Fig. \ref{fig:ViiiTEMs} (b), (d)). The change of nanowire structure from zincblende to wurtzite with increasing V/III has been reported experimentally\cite{lehmann2015crystal, jacobsson_interface_2016} and theoretically\cite{dubrovskii_mono-_2015}. MOCVD experiments also show a second transition back to zincblende at very high V/III ratio\cite{lehmann2015crystal,maartensson2019simulation} but this regime was not covered in the experiments described here. We have also observed in other experiments that, for even higher TMGa flows than studied here, the interface develops an oscillating truncation consistent with earlier \textit{in situ} observations.\cite{wen_periodically_2011,gamalski2011cyclic,chou_atomic-scale_2014,jacobsson_interface_2016} Since the interface dynamics are qualitatively very different in that regime, it was not covered during the experiments that are included in this study.

\section*{Conclusions}
In summary, GaAs nanowires were grown with a gold-based catalyst particle in an environmental TEM in order to deepen the understanding of nanowire growth. The chemical composition of the catalyst particle was measured \textit{in situ} as the nanowire was growing. We report the catalyst composition during growth of one nanowire as a function of temperature and another wire of similar dimensions as a function of ratio of precursor flux. The Ga content in the catalyst increases with increasing temperature which correlates well with calculated phase diagrams. Since the concentration seems to be determined by thermodynamics, the results would be applicable to Au-catalyzed growth of GaAs nanowires independent of the growth method. Although the As content in the catalyst is below the detection limit by XEDS, by comparing measured Ga-Au content with calculated Au-Ga-As phase diagrams we can estimate the As concentration. The Ga concentration in the catalyst also increases with increasing Ga-precursor flux. These \textit{in situ} measurements will aid better theoretical modelling of nanowire growth and improve the understanding of nanowire growth mechanisms. Most metal assisted III-V and II-VI nanowire growths typically have low solubility of the anion in the catalyst and the Au assisted GaAs studied here serves as a model system.

\section*{Methods}
Au aerosol particles of 30 nm diameter on average were used to seed nanowire growth. A silicon nitride MEMS heating chip from Norcada was used as substrate. The thinnest SiN$_x$ parts where growth was monitored had a thickness of 30 nm. Atomic resolved imaging was performed with an AMT XR401 sCMOS camera and the videos were recorded at about 20 fps. The TEM images in the article were extracted from these videos and processed.

GaAs nanowires were grown in a Hitachi HF3300S environmental transmission electron microscope (ETEM) with CEOS B-COR-aberration-corrector and a cold field emission gun. Blaze software by Hitachi was used to control the local sample temperature using Joule heating in a constant power mode. The ETEM was connected to a gas handling system with the CVD gases. A single tilt holder that has two separate microtubes running to the holder tip was used for supplying gases. The holder and the gas handling system are connected by a polymer coated thin quartz tube (PEEKSil) from Trajan Scientific. 

Pressure at the sample: During growth experiments the pressure near the pole piece was measured by an Inficon MPG400 pressure gauge and is referred to as `column pressure' here. The precursor inlet tubes run along the length of the TEM holder, and precursor gases are released close to the heated SiN$_x$ grid. Hence the pressures are higher at the growth front than the `column pressure'. The sample pressure relative to the column pressure was calibrated using the pressure at the heating coil of a clean SiN$_x$ grid (without Au or GaAs) as a pressure gauge following the Pirani gauge principles using the Blaze software. We performed a calibration experiments with N$_2$ and also H$_2$ and found that the pressure at the sample (measured by Blaze) is twice of `column pressure'.  A factor of 2 is therefore used to estimate sample pressure for each species based on its calibrated column pressure.

Precursor partial pressures and V/III ratio:  The TMGa bubbler was maintained at -10~$^\circ$C with H$_2$ bubbled through it. Additionally, a small fixed amount of H$_2$ dilution (to be more precise, four times the flow used for the bubbler) was added in the TMGa line followed the bubbler. No additional carrier gas was used, and the H$_2$ partial pressure is thus much lower than in a typical MOCVD. The flow of the TMGa/H$_2$ mixture was controlled by mass flow controller (MFC), and a portion of the resulting flow was bypassed to the vent line to restrict the pressure reaching the microscope. 

In order to determine the partial pressure of TMGa at the sample, the precursor fluxes sent to the ETEM were monitored with a residual gas analyzer (SRS RGA 300) using mass spectrometry in these experiments. The amount of the dominant TMGa derivatives (containing Ga) are measured at mass-to-charge ratio of 101, 99, 71 and 69 associated with dimethylgallium and Ga. The sample heating is very local at the SiN$_x$ grid, decomposing just a very small fraction of the supplied precursors and so these RGA measurements are independent of localized pyrolysis at the sample.
Calibration experiments were performed for different but known TMGa and H$_2$ flows to find the correlation between `column pressure' and the sum of the Ga-related mass spectrometry readings (`column pressure' in this case was corrected for H$_2$ using Inficon MPG400 pressure gauge calibration factor). TMGa partial pressure at the sample during experiments is therefore determined using the calibrated RGA readings together with the factor of 2 between column pressure and sample pressure described above.

The AsH$_3$ flow was controlled exclusively by MFC (no part bypassed) and fixed for all experiments to give an estimated pressure of 0.1 Pa `column pressure', i.e. equivalently 0.2 Pa at the sample. Since the Inficon MPG400 pressure gauge is originally calibrated for N$_2$, this AsH$_3$ `column pressure' was measured in comparison to an equivalent flow of N$_2$ gas. V/III ratio is then calculated using this value divided by the TMGa pressure calculated using RGA readings for the specific experiment.	

XEDS : The XEDS measurement was performed with a SDD X-Max$^N$ 80 T system from Oxford Instruments.  While measuring the catalyst composition a small condenser aperture was used and condensed the beam on the anterior part of the catalyst (opposite to the nanowire/catalyst interface). As the nanowire grows the catalyst particle keeps moving forward. During XEDS the illuminated area on the fluorescent screen was continuously monitored and the sample moved appropriately. We specify the percentage content of the elements in terms of atomic percentage (and not as weight percentage) throughout this paper. Data was acquired and quantified with Aztec software. Acquisition was typically for 4 minutes. In XEDS scans we observe signals of Au, Ga and As from the catalyst and/or nanowire, Si and N from the substrate and Cu due to scattering from parts of the microscope. The sum of Ga and Au signal is normalized to 100 atomic~\% to obtain results quoted in this article. The default lines (K for Ga, L for Au, K for As) were used for quantification. The standard deviation of quantification result is shown as the y-axis error in Fig. \ref{fig:temp} (b) and Fig. \ref{fig:VIIIedx} (b). More details about the XEDS quantification results are given in Supplementary Information section S3.

Arsenic content from XEDS: Typically, the measured As signal is very weak and a clear peak cannot necessarily be distinguished from background. Full quantification of spectra (including background and artefacts such as Si, N, etc) yields an As weight~\% of less than one, which is close to the detection limit in XEDS. Moreover, any small As signal detected may not necessarily originate from the catalyst nanoparticle. Since the nanowire is growing during the XEDS acquisition, the catalyst/nanowire interface moves; although spectra were acquired carefully to avoid the electron beam interacting with the nanowire directly, electrons scattered into the nanowire from the catalyst could easily lead to an overestimate of both Ga and As by a few percent.  

Temperature series experiment: In the temperature series experiment discussed, temperature was increased from 440~$^\circ$C to 500~$^\circ$C in steps of 20~$^\circ$C, and finally decreased to 420~$^\circ$C. After each temperature was reached we waited at least a minute so that the catalyst stabilizes and there is no evident change in its dimensions. This wire grew in the wurtzite structure in the $<$0001$>$ direction at the conditions studied. 

V/III series experiment: The measurement for the V/III series were recorded at 420~$^\circ$C. We started to observe the nanowire when the V/III was 1535 where we measured the first EDX data. Then the Ga supply was stopped for some time. The TMGa flow was restarted and increased slowly in steps until (at very high flow) the nanowire changed direction and folded back on itself during an XEDS acquisition. 
During the XEDS measurement at the lowest V/III ratio (179) there was some As signal observed due to scattering from GaAs NW, so the same percentage of Ga was subtracted from the quantification results and renormalized to obtain the data point plotted. (See Supplementary data section S3 for details)
The range about which V/III varied during individual XEDS spectrum acquisition is denoted by the error bar for the x-axis in Fig. \ref{fig:VIIIedx} (b).

\section*{Data Availability} More data that support the findings of this study are available from the corresponding author upon request.

\section*{References}


\begin{thebibliography}{10}

\bibitem{guo2006structural}
Y.~N. Guo, J.~Zou, M.~Paladugu, H.~Wang, Q.~Gao, H.~H. Tan, and C.~Jagadish.
\newblock Structural characteristics of {GaSb/GaAs} nanowire heterostructures
  grown by metal-organic chemical vapor deposition.
\newblock {\em Applied physics letters}, 89(23):231917, 2006.

\bibitem{larsson2006strain}
M.~W. Larsson, J.~B. Wagner, M.~Wallin, P.~H{\aa}kansson, L.~E. Fr{\"o}berg,
  L.~Samuelson, and L.~R. Wallenberg.
\newblock Strain mapping in free-standing heterostructured wurtzite {InAs/InP}
  nanowires.
\newblock {\em Nanotechnology}, 18(1):015504, 2006.

\bibitem{caroff2009insb}
P.~Caroff, M.~E. Messing, B.~M. Borg, K.~A. Dick, K.~Deppert, and L.~E.
  Wernersson.
\newblock {InSb} heterostructure nanowires: {MOVPE} growth under extreme
  lattice mismatch.
\newblock {\em Nanotechnology}, 20(49):495606, 2009.

\bibitem{persson_solid-phase_2004}
A.~I. Persson, M.~W. Larsson, S.~Stenstr{\"o}m, B.~J. Ohlsson, L.~Samuelson,
  and L.~R. Wallenberg.
\newblock Solid-phase diffusion mechanism for {GaAs} nanowire growth.
\newblock {\em Nature Materials}, 3(10):677--681, October 2004.

\bibitem{jacobs2007electronic}
B.~W. Jacobs, V.~M. Ayres, M.~P. Petkov, J.~B. Halpern, M.~He, A.~D. Baczewski,
  K.~McElroy, M.~A. Crimp, J.~Zhang, and H.~C. Shaw.
\newblock Electronic and structural characteristics of zinc-blende wurtzite
  biphasic homostructure {GaN} nanowires.
\newblock {\em Nano letters}, 7(5):1435--1438, 2007.

\bibitem{joyce_phase_2010}
H.~J. Joyce, J.~Wong-Leung, Q.~Gao, H.~H. Tan, and C.~Jagadish.
\newblock Phase perfection in zinc blende and wurtzite {III}-{V} nanowires
  using basic growth parameters.
\newblock {\em Nano Letters}, 10(3):908--915, March 2010.

\bibitem{dick_control_2010}
K.~A. Dick, P.~Caroff, J.~Bolinsson, M.~E. Messing, J.~Johansson, K.~Deppert,
  L.~R. Wallenberg, and L.~Samuelson.
\newblock Control of {III}-{V} nanowire crystal structure by growth parameter
  tuning.
\newblock {\em Semiconductor Science and Technology}, 25(2):024009, 2010.

\bibitem{sukrittanon2014growth}
S.~Sukrittanon, Y.~J. Kuang, A.~Dobrovolsky, W.~M. Kang, J.~S. Jang, B.~J. Kim,
  W.~M. Chen, I.~A. Buyanova, and C.~W. Tu.
\newblock Growth and characterization of dilute nitride {GaN$_{x}$P$_{1-x}$}
  nanowires and {GaN$_{x}$P$_{1-x}$/GaN$_{y}$P$_{1-y}$} core/shell nanowires on
  {Si} (111) by gas source molecular beam epitaxy.
\newblock {\em Applied Physics Letters}, 105(7):072107, 2014.

\bibitem{namazi2017direct}
L.~Namazi, S.~G. Ghalamestani, S.~Lehmann, R.~R. Zamani, and K.~A. Dick.
\newblock Direct nucleation, morphology and compositional tuning of {InAs$_{1-
  x}$Sb$_{x}$} nanowires on {InAs} (111) {B} substrates.
\newblock {\em Nanotechnology}, 28(16):165601, 2017.

\bibitem{dubrovskii_refinement_2017}
V.~G. Dubrovskii.
\newblock Refinement of nucleation theory for vapor–liquid–solid nanowires.
\newblock {\em Crystal Growth \& Design}, 17(5):2589--2593, May 2017.

\bibitem{schwarz_elementary_2011}
K.~W. Schwarz and J.~Tersoff.
\newblock Elementary processes in nanowire growth.
\newblock {\em Nano Letters}, 11(2):316--320, February 2011.

\bibitem{nebolsin_role_2003}
V.~A. Nebol\textsc{\char13}sin and A.~A. Shchetinin.
\newblock Role of surface energy in the vapor--liquid--solid growth of silicon.
\newblock {\em Inorganic Materials}, 39(9):899--903, September 2003.

\bibitem{krogstrup_advances_2013}
P.~Krogstrup, H.~I. J{\o}rgensen, E.~Johnson, M.~H. Madsen, Claus~B.
  S{\o}rensen, A.~F. Morral, M.~Aagesen, J.~Nyg{\aa}rd, and F.~Glas.
\newblock Advances in the theory of {III}-{V} nanowire growth dynamics.
\newblock {\em Journal of Physics D: Applied Physics}, 46(31):313001, 2013.

\bibitem{glas_chemical_2010}
F.~Glas.
\newblock Chemical potentials for {Au}-assisted vapor-liquid-solid growth of
  {III}-{V} nanowires.
\newblock {\em Journal of Applied Physics}, 108(7):073506, October 2010.

\bibitem{dubrovskii_group_2016}
V.~G. Dubrovskii.
\newblock Group {V} sensitive vapor--liquid--solid growth of {Au}--catalyzed
  and self--catalyzed {III}-{V} nanowires.
\newblock {\em Journal of Crystal Growth}, 440:62--68, April 2016.

\bibitem{plante_analytical_2009}
M.~C. Plante and R.~R. LaPierre.
\newblock Analytical description of the metal-assisted growth of {III}-{V}
  nanowires: Axial and radial growths.
\newblock {\em Journal of Applied Physics}, 105(11):114304, June 2009.

\bibitem{ramdani_arsenic_2013}
M.~R. Ramdani, J.~C. Harmand, F.~Glas, G.~Patriarche, and L.~Travers.
\newblock Arsenic pathways in self-catalyzed growth of {GaAs} nanowires.
\newblock {\em Crystal Growth \& Design}, 13(1):91--96, January 2013.

\bibitem{dubrovskii_diffusion-controlled_2007}
V.~G. Dubrovskii, N.~V. Sibirev, R.~A. Suris, G.~E. Cirlin, J.~C. Harmand, and
  V.~M. Ustinov.
\newblock Diffusion--controlled growth of semiconductor nanowires: Vapor
  pressure versus high vacuum deposition.
\newblock {\em Surface Science}, 601(18):4395--4401, September 2007.

\bibitem{glas_why_2007}
F.~Glas, J.~C. Harmand, and G.~Patriarche.
\newblock Why does wurtzite form in nanowires of {III}-{V} zinc blende
  semiconductors?
\newblock {\em Physical Review Letters}, 99(14):146101, October 2007.

\bibitem{dubrovskii_growth_2008}
V.~G. Dubrovskii, N.~V. Sibirev, J.~C. Harmand, and F.~Glas.
\newblock Growth kinetics and crystal structure of semiconductor nanowires.
\newblock {\em Physical Review B}, 78(23):235301, December 2008.

\bibitem{johansson2015polytype}
J.~Johansson, Z.~Zanolli, and K.~A. Dick.
\newblock Polytype attainability in {III}-{V} semiconductor nanowires.
\newblock {\em Crystal growth \& design}, 16(1):371--379, 2015.

\bibitem{tornberg2017thermodynamic}
M.~Tornberg, K.~A. Dick, and S.~Lehmann.
\newblock Thermodynamic stability of gold-assisted {InAs} nanowire growth.
\newblock {\em The Journal of Physical Chemistry C}, 121(39):21678--21684,
  2017.

\bibitem{panciera_controlling_2016}
F.~Panciera, M.~M. Norton, S.~B. Alam, S.~Hofmann, K.~M{\o}lhave, and F.~M.
  Ross.
\newblock Controlling nanowire growth through electric field-induced
  deformation of the catalyst droplet.
\newblock {\em Nature Communications}, 7:12271, July 2016.

\bibitem{dubrovskii_mono-_2015}
V.~G. Dubrovskii.
\newblock Mono- and polynucleation, atomistic growth, and crystal phase of
  {III}-{V} nanowires under varying group {V} flow.
\newblock {\em The Journal of Chemical Physics}, 142(20):204702, May 2015.

\bibitem{krogstrup_impact_2011}
P.~Krogstrup, S.~Curiotto, E.~Johnson, M.~Aagesen, J.~Nyg{\aa}rd, and
  D.~Chatain.
\newblock Impact of the liquid phase shape on the structure of {III}-{V}
  nanowires.
\newblock {\em Physical Review Letters}, 106(12):125505, March 2011.

\bibitem{algra_formation_2011}
R.~E. Algra, V.~Vonk, D.~Wermeille, W.~J. Szweryn, M.~A. Verheijen, W.~J.~P.
  van Enckevort, A.~A.~C. Bode, W.~L. Noorduin, E.~Tancini, A.~E.~F. de~Jong,
  E.~P. A.~M. Bakkers, and E.~Vlieg.
\newblock Formation of wurtzite {InP} nanowires explained by liquid-ordering.
\newblock {\em Nano Letters}, 11(1):44--48, January 2011.

\bibitem{dubrovskii_general_2007}
V.~G. Dubrovskii and N.~V. Sibirev.
\newblock General form of the dependences of nanowire growth rate on the
  nanowire radius.
\newblock {\em Journal of crystal growth}, 304(2):504--513, 2007.

\bibitem{froberg2007diameter}
L.~E. Fr{\"o}berg, W.~Seifert, and J.~Johansson.
\newblock Diameter-dependent growth rate of {InAs} nanowires.
\newblock {\em Physical Review B}, 76(15):153401, 2007.

\bibitem{li2013nucleation}
N.~Li, W.~Li, L.~Liu, and T.~Y. Tan.
\newblock A nucleation-growth model of nanowires produced by the
  vapor-liquid-solid process.
\newblock {\em Journal of Applied Physics}, 114(6):064302, 2013.

\bibitem{borgstrom2007synergetic}
M.~T. Borgstr{\"o}m, G.~Immink, B.~Ketelaars, R.~Algra, and E.~P. A.~M.
  Bakkers.
\newblock Synergetic nanowire growth.
\newblock {\em Nature nanotechnology}, 2(9):541, 2007.

\bibitem{li2013readsorption}
A.~Li, N.~V. Sibirev, D.~Ercolani, V.~G Dubrovskii, and L.~Sorba.
\newblock Readsorption assisted growth of {InAs/InSb} heterostructured nanowire
  arrays.
\newblock {\em Crystal Growth \& Design}, 13(2):878--882, 2013.

\bibitem{borg2013geometric}
B.~M. Borg, J.~Johansson, K.~Storm, and K.~Deppert.
\newblock Geometric model for metalorganic vapour phase epitaxy of dense
  nanowire arrays.
\newblock {\em Journal of Crystal Growth}, 366:15--19, 2013.

\bibitem{ek_atomic-scale_2018}
M.~Ek and M.~A. Filler.
\newblock Atomic--scale choreography of vapor--liquid--solid nanowire growth.
\newblock {\em Accounts of Chemical Research}, 51(1):118--126, January 2018.

\bibitem{kirkham_tracking_2010}
M.~Kirkham, Z.~L. Wang, and R.~L. Snyder.
\newblock Tracking the catalyzed growth process of nanowires by in situ x-ray
  diffraction.
\newblock {\em Journal of Applied Physics}, 108(1):014304, 2010.

\bibitem{schroth_evolution_2015}
P.~Schroth, M.~K{\"o}hl, J.~W. Hornung, E.~Dimakis, C.~Somaschini, L.~Geelhaar,
  A.~Biermanns, S.~Bauer, S.~Lazarev, U.~Pietsch, et~al.
\newblock Evolution of polytypism in {GaAs} nanowires during growth revealed by
  time-resolved in situ x-ray diffraction.
\newblock {\em Physical review letters}, 114(5):055504, 2015.

\bibitem{krogstrup_situ_2012}
P.~Krogstrup, M.~Hannibal~Madsen, W.~Hu, M.~Kozu, Y.~Nakata, J.~Nyg{\aa}rd,
  M.~Takahasi, and R.~Feidenhans\textsc{\char13}l.
\newblock In-situ x-ray characterization of wurtzite formation in {GaAs}
  nanowires.
\newblock {\em Applied Physics Letters}, 100(9):093103, 2012.

\bibitem{sivaram_surface_2016}
S.~V. Sivaram, H.~Y. Hui, M.~de~la Mata, J.~Arbiol, and M.~A. Filler.
\newblock Surface hydrogen enables subeutectic vapor--liquid--solid
  semiconductor nanowire growth.
\newblock {\em Nano Letters}, 16(11):6717--6723, November 2016.

\bibitem{sivaram2015direct}
S.~V. Sivaram, N.~Shin, L.~W. Chou, and M.~A. Filler.
\newblock Direct observation of transient surface species during {Ge} nanowire
  growth and their influence on growth stability.
\newblock {\em Journal of the American Chemical Society}, 137(31):9861--9869,
  2015.

\bibitem{shin2014interplay}
N.~Shin, M.~Chi, and M.~A. Filler.
\newblock Interplay between defect propagation and surface hydrogen in silicon
  nanowire kinking superstructures.
\newblock {\em ACS nano}, 8(4):3829--3835, 2014.

\bibitem{tchernycheva2006temperature}
M.~Tchernycheva, J.~C. Harmand, G.~Patriarche, L.~Travers, and G.~E. Cirlin.
\newblock Temperature conditions for {GaAs} nanowire formation by {Au}-assisted
  molecular beam epitaxy.
\newblock {\em Nanotechnology}, 17(16):4025, 2006.

\bibitem{xu2012faceting}
T.~Xu, K.~A. Dick, S.~Plissard, T.~H. Nguyen, Y.~Makoudi, M.~Berthe, J.~P. Nys,
  X.~Wallart, B.~Grandidier, and P.~Caroff.
\newblock Faceting, composition and crystal phase evolution in {III}-{V}
  antimonide nanowire heterostructures revealed by combining microscopy
  techniques.
\newblock {\em Nanotechnology}, 23(9):095702, 2012.

\bibitem{jo2018real}
J.~Jo, Y.~Tchoe, G.~C. Yi, and M.~Kim.
\newblock Real-time characterization using in situ {RHEED} transmission mode
  and {TEM} for investigation of the growth behaviour of nanomaterials.
\newblock {\em Scientific reports}, 8(1):1694, 2018.

\bibitem{heurlin_situ_2015}
M.~Heurlin, N.~Anttu, C.~Camus, L.~Samuelson, and M.~T. Borgstr{\"o}m.
\newblock In situ characterization of nanowire dimensions and growth dynamics
  by optical reflectance.
\newblock {\em Nano letters}, 15(5):3597--3602, 2015.

\bibitem{clement2006situ}
T.~Clement, S.~Ingole, S.~Ketharanathan, J.~Drucker, and S.~T. Picraux.
\newblock In situ studies of semiconductor nanowire growth using optical
  reflectometry.
\newblock {\em Applied physics letters}, 89(16):163125, 2006.

\bibitem{fernandez_monitoring_2015}
S.~Fern{\'a}ndez-Garrido, J.~K. Zettler, L.~Geelhaar, and O.~Brandt.
\newblock Monitoring the formation of nanowires by line-of-sight quadrupole
  mass spectrometry: a comprehensive description of the temporal evolution of
  gan nanowire ensembles.
\newblock {\em Nano letters}, 15(3):1930--1937, 2015.

\bibitem{kolibal_synergic_2016}
M.~Kol{\'\i}bal, T.~Pejchal, T.~Vystavel, and T.~Sikola.
\newblock The synergic effect of atomic hydrogen adsorption and catalyst
  spreading on {Ge} nanowire growth orientation and kinking.
\newblock {\em Nano letters}, 16(8):4880--4886, 2016.

\bibitem{kodambaka_germanium_2007}
S.~Kodambaka, J.~Tersoff, M.~C. Reuter, and F.~M. Ross.
\newblock Germanium nanowire growth below the eutectic temperature.
\newblock {\em Science}, 316(5825):729--732, May 2007.

\bibitem{hofmann_ledge-flow-controlled_2008}
S.~Hofmann, R.~Sharma, C.~T. Wirth, F.~Cervantes-Sodi, C.~Ducati, T.~Kasama,
  R.~E. Dunin-Borkowski, J.~Drucker, P.~Bennett, and J.~Robertson.
\newblock Ledge-flow-controlled catalyst interface dynamics during {Si}
  nanowire growth.
\newblock {\em Nature Materials}, 7(5):372, May 2008.

\bibitem{wen_periodically_2011}
C.-Y. Wen, J.~Tersoff, K.~Hillerich, M.~C. Reuter, J.~H. Park, S.~Kodambaka,
  E.~A. Stach, and F.~M. Ross.
\newblock Periodically changing morphology of the growth interface in {Si},
  {Ge}, and {GaP} nanowires.
\newblock {\em Physical Review Letters}, 107(2):025503, July 2011.

\bibitem{gamalski2011cyclic}
A.~D. Gamalski, C.~Ducati, and S.~Hofmann.
\newblock Cyclic supersaturation and triple phase boundary dynamics in
  germanium nanowire growth.
\newblock {\em The Journal of Physical Chemistry C}, 115(11):4413--4417, 2011.

\bibitem{chou_atomic-scale_2014}
Y.-C. Chou, K.~Hillerich, J.~Tersoff, M.~C. Reuter, K.~A. Dick, and F.~M. Ross.
\newblock Atomic-scale variability and control of {III}-{V} nanowire growth
  kinetics.
\newblock {\em Science}, 343(6168):281--284, January 2014.

\bibitem{jacobsson_interface_2016}
D.~Jacobsson, F.~Panciera, J.~Tersoff, M.~C. Reuter, S.~Lehmann, S.~Hofmann,
  K.~A. Dick, and F.~M. Ross.
\newblock Interface dynamics and crystal phase switching in {GaAs} nanowires.
\newblock {\em Nature}, 531(7594):317, March 2016.

\bibitem{harmand_atomic_2018}
J.~C. Harmand, G.~Patriarche, F.~Glas, F.~Panciera, I.~Florea, J.~L. Maurice,
  L.~Travers, and Y.~Ollivier.
\newblock Atomic step flow on a nanofacet.
\newblock {\em Physical Review Letters}, 121(16):166101, October 2018.

\bibitem{gamalski_atomic_2016}
A.~D. Gamalski, J.~Tersoff, and E.~A. Stach.
\newblock Atomic resolution in situ imaging of a double-bilayer multistep
  growth mode in gallium nitride nanowires.
\newblock {\em Nano Letters}, 16(4):2283--2288, April 2016.

\bibitem{dubrovskii_composition_2018}
V.~G. Dubrovskii, Z.~V. Sokolova, M.~V. Rylkova, and A.~A. Zhiglinsky.
\newblock Composition and contact angle of {Au}-{III}-{V} droplets on top of
  {Au}-catalyzed {III}-{V} nanowires.
\newblock {\em Materials Physics and Mechanics}, 36(1):1--7, 2018.

\bibitem{dubrovskii_zeldovich_2015}
V.~G. Dubrovskii and J.~Grecenkov.
\newblock Zeldovich nucleation rate, self-consistency renormalization, and
  crystal phase of {Au}-catalyzed {GaAs} nanowires.
\newblock {\em Crystal Growth \& Design}, 15(1):340--347, January 2015.

\bibitem{dubrovskii_influence_2014}
V.~G. Dubrovskii.
\newblock Influence of the group {V} element on the chemical potential and
  crystal structure of {Au}-catalyzed {III}-{V} nanowires.
\newblock {\em Applied Physics Letters}, 104(5):053110, February 2014.

\bibitem{leshchenko_nucleation-limited_2018}
E.~D. Leshchenko, M.~Ghasemi, V.~G. Dubrovskii, and J.~Johansson.
\newblock Nucleation-limited composition of ternary {III}-{V} nanowires forming
  from quaternary gold based liquid alloys.
\newblock {\em CrystEngComm}, 20(12):1649--1655, 2018.

\bibitem{harmand_analysis_2005}
J.~C. Harmand, G.~Patriarche, N.~P{\'e}r{\'e}-Laperne, M.~N. M{\'e}rat-Combes,
  L.~Travers, and F.~Glas.
\newblock Analysis of vapor-liquid-solid mechanism in {Au}-assisted {GaAs}
  nanowire growth.
\newblock {\em Applied Physics Letters}, 87(20):203101, November 2005.

\bibitem{jacobsson_zincblende--wurtzite_2013}
D.~Jacobsson, S.~Lehmann, and K.~A. Dick.
\newblock Zincblende-to-wurtzite interface improvement by group {III} loading
  in {Au}-seeded {GaAs} nanowires.
\newblock {\em physica status solidi (RRL) – Rapid Research Letters},
  7(10):855--859, October 2013.

\bibitem{chatillon_thermodynamics_2009}
C.~Chatillon, F.~Hodaj, and A.~Pisch.
\newblock Thermodynamics of {GaAs} nanowire {MBE} growth with gold droplets.
\newblock {\em Journal of Crystal Growth}, 311(14):3598--3608, July 2009.

\bibitem{tersoff_stable_2015}
J.~Tersoff.
\newblock Stable self-catalyzed growth of {III}-{V} nanowires.
\newblock {\em Nano Letters}, 15(10):6609--6613, October 2015.

\bibitem{tizei_iiiv_2009}
L.~H.~G. Tizei, T.~Chiaramonte, D.~Ugarte, and M.~A. Cotta.
\newblock {III}-{V} semiconductor nanowire growth: does arsenic diffuse through
  the metal nanoparticle catalyst.
\newblock {\em Nanotechnology}, 20(27):275604, 2009.

\bibitem{glas2013predictive}
F.~Glas, M.~R. Ramdani, G.~Patriarche, and J.~C. Harmand.
\newblock Predictive modeling of self-catalyzed {III}-{V} nanowire growth.
\newblock {\em Physical Review B}, 88(19):195304, 2013.

\bibitem{maartensson2019simulation}
E.~K. M{\aa}rtensson, S.~Lehmann, K.~A. Dick, and J.~Johansson.
\newblock Simulation of {GaAs} nanowire growth and crystal structure.
\newblock {\em Nano letters}, 2019.

\bibitem{ghasemi_phase_2017}
M.~Ghasemi and J.~Johansson.
\newblock Phase diagrams for understanding gold-seeded growth of {GaAs} and
  {InAs} nanowires.
\newblock {\em Journal of Physics D: Applied Physics}, 50(13):134002, 2017.

\bibitem{lehmann2015crystal}
S.~Lehmann, D.~Jacobsson, and K.~A. Dick.
\newblock Crystal phase control in {GaAs} nanowires: opposing trends in the
  {Ga}-and {As}-limited growth regimes.
\newblock {\em Nanotechnology}, 26(30):301001, 2015.

\bibitem{dheeraj2012controlling}
D.~L. Dheeraj, A.~M. Munshi, M.~Scheffler, A.T.~J. van Helvoort, H.~Weman, and
  B.~O. Fimland.
\newblock Controlling crystal phases in {GaAs} nanowires grown by {Au}-assisted
  molecular beam epitaxy.
\newblock {\em Nanotechnology}, 24(1):015601, 2012.

\bibitem{dubrovskii2018composition}
V.~G. Dubrovskii, Z.~V. Sokolova, M.~V. Rylkova, and A.~A. Zhiglinsky.
\newblock Composition and contact angle of {Au-III-V} droplets on top of {Au-C}
  {III-V} nanowires.
\newblock {\em Materials Physics and Mechanics}, 36:1--7, 2018.

\bibitem{hannon2006influence}
J.~B. Hannon, S.~Kodambaka, F.~M. Ross, and R.~M. Tromp.
\newblock The influence of the surface migration of gold on the growth of
  silicon nanowires.
\newblock {\em Nature}, 440(7080):69, 2006.

\bibitem{otnes2016strategies}
G.~Otnes, M.~Heurlin, M.~Graczyk, J.~Wallentin, D.~Jacobsson, A.~Berg,
  I.~Maximov, and M.~T. Borgstr{\"o}m.
\newblock Strategies to obtain pattern fidelity in nanowire growth from
  large-area surfaces patterned using nanoimprint lithography.
\newblock {\em Nano Research}, 9(10):2852--2861, 2016.

\end{thebibliography}

\bibliographystyle{unsrt}



\section*{Acknowledgments}
We wish to acknowledge support from the Knut and Alice Wallenberg Foundation, NanoLund and the Swedish Research Council. We thank E. M\aa rtensson for valuable discussions and insights. 
We thank Joacim Gustafsson, Stas Dogel and Charles Soong from Hitachi and John Wheeler and Shaun Ohman from Collabratech for technical assistance in setting up the laboratory.

\section*{Author contributions}
CBM, DJ, MT and ARP performed the experiments. Data analysis was done by CBM. JJ performed the phase diagram calculations. KAD and RW coordinated the project. All authors discussed the results and contributed to the concepts discussed in this article. 

\section*{Competing financial interests}
The authors declare no competing financial interests.


\newpage

\setcounter{figure}{0}
\makeatletter
\renewcommand{\thefigure}{S\arabic{figure}}
\renewcommand\thesection{\ S\arabic {section}}
\renewcommand\thetable{\ S\arabic {table}}



\newpage

\section*{\large {Supplementary Information: \textit{In situ} analysis of catalyst composition during gold catalyzed GaAs nanowire growth}}
\vspace{1cm}

\vspace{1cm}

\section{Examples of catalyst composition measured \textit{ex situ} from Au-assisted GaAs nanowire}
The catalyst composition measured post-growth depends on the conditions used to terminate the growth.  
Some representative examples are shown in Table S1.  
\begin{table}[!ht]
\centering
\begin{tabular}{|c|p{2.cm}|p{8.3cm}|p{0.8cm}|}
\hline
\textbf{Reference} & \textbf{Growth} \qquad \textbf{Technique} & \textbf{Cooling / growth termination} & \textbf{Ga\%}   \\ \hline
 &  & Cooling in H$_2$ environment of zincblende nanowires (grown with relatively less Ga precursor flux) & 21 \\ \cline{3-4}
Jacobsson \textit{et al.}\cite{jacobsson_zincblende--wurtzite_2013} & MOCVD & Cooling in AsH$_3$/H$_2$ environment of zincblende nanowires (grown with relatively less Ga precursor flux) & $<$3  \\ \cline{3-4}
& & Cooling in H$_2$ environment of wurtzite nanowires (grown with relatively larger Ga precursor flux) & 31  \\ \cline{3-4}
&  & Cooling in AsH$_3$/H$_2$ environment of wurtzite nanowires (grown with relatively larger Ga precursor flux) & $\sim$3  \\ \hline
 &  & Ga stopped before As$_2$ after growth & 1 \\ \cline{3-4}
Harmand \textit{et al.}\cite{harmand_analysis_2005} & Molecular Beam & Ga and As$_2$  stopped simultaneously after growth & 50  \\ \cline{3-4}
& Epitaxy & Longer growth time to get tapered nanowires, then Ga and As$_2$ stopped simultaneously & 31  \\ \hline
 
Persson \textit{et al.}\cite{persson_solid-phase_2004} & Chemical Beam & Ga and As precursors stopped simultaneously after growth & 9 \\ \cline{3-4}
&Epitaxy & Cooled down in As & 0 \\ \hline
\end{tabular}
\caption{Ga content in the catalyst measured \textit{ex situ} after growth of Au-seeded GaAs nanowires as reported in literature. The composition depends strongly on the growth and cooling conditions.
}
\label{table1}
\end{table}

\section{Statistics: XEDS from different nanowires at same conditions}

It is important to know that the catalyst composition across different particles is roughly the same at identical conditions. Hence we conducted experiment at 420~$^\circ$C with a fixed low TMGa flow and measured the Ga concentration across different nanowire catalysts. They had roughly the same concentration. We also looked at diameter dependence of catalyst composition and found little difference in the range we measured (16 nm to 66 nm diameter during growth) 
(For this experiment we used Au nanoparticles with different sizes unlike the two sets of experiments reported in the article.)

\begin{figure}
\centerline {\includegraphics[width=88mm]{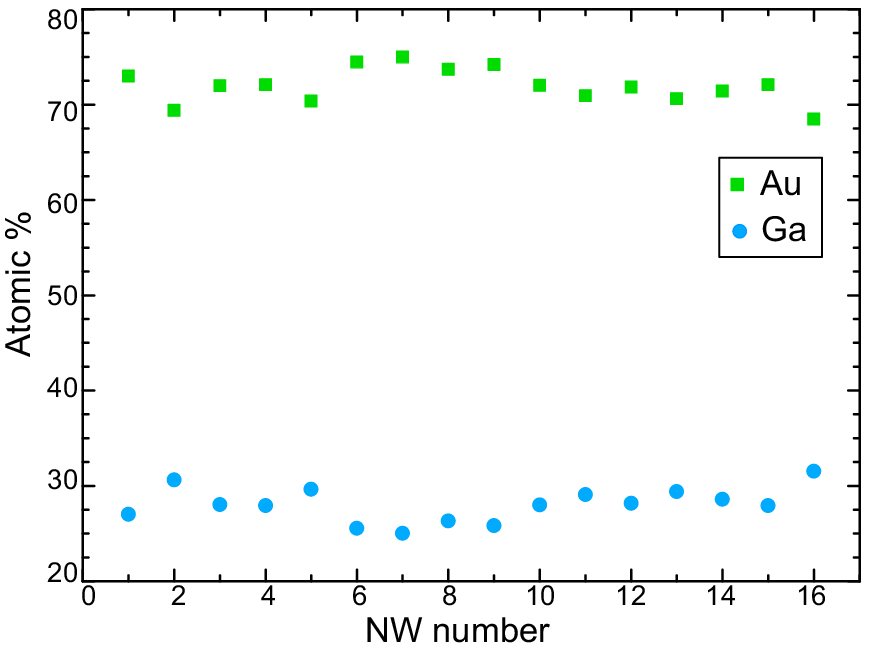}}
\caption{ \label{fig:edxstat} The Ga and Au \% of different catalyst particles measured \textit{in situ} at identical growth conditions (V/III=1500, 420 $^\circ$C). }
\end{figure}	

\newpage
\section{XEDS data measured for the temperature series and the V/III series}

In the article (Fig. 2 (b) and 5 (b)) only the Ga concentration in the Au-Ga-As catalyst alloy was plotted. The XEDS quantification data measured including the background signal from the SiN$_x$ film is shown in the tables below. Table \ref{tabletemp} is of the temperature series shown in Fig. 2 (b). Table \ref{tableVIII} is of the V/III series shown in Fig. 5 (b). 
We suspect the As\% observed arises mainly due to scattering from the adjacent GaAs nanowire. 
(Quantification of low energy peaks $<$0.5 keV are not available at 480 and 500 $^\circ$C due to drastically increased strobe peak.)

\begin{table}[!h]
\centering
\begin{tabular}{|c|c|c|c|c|c|c|}
\hline
\textbf{T($^\circ$C)} & \textbf{Ga }& \textbf{As} & \textbf{Au} & \textbf{N} & \textbf{Si} & \textbf{Cu} \\ \hline
440	& 3.5	& 0.2	& 8.0	  & 45.1	& 40.3	& 2.9 \\ 
460	& 5.5	& 0.4	& 11.4	& 25.8	& 52.7	& 4.2 \\ 
480	& 9.2	& 0.7	& 16.8	& -	  & 67.3	& 6.0 \\ 
500	& 9.6	& 0.8	& 16.5	& -	  & 67.4	& 5.9  \\ 
420	& 4.6 &	0.6	& 13.5	& 38.8	& 38.0	& 4.4 \\ \hline
\end{tabular}
\caption{The XEDS quantification results for all the elements present corresponding to the temperature series measurement plotted in Fig. 2 (b).}
\label{tabletemp}
\end{table}

\begin{table}[!h]
\centering
\begin{tabular}{|c|c|c|c|c|c|c|}
\hline
\textbf{V/III}	& \textbf{Ga }& \textbf{As} & \textbf{Au} & \textbf{N} & \textbf{Si} & \textbf{Cu}  \\ \hline
1535 & 4.6	& 0.3	& 9.7	& 45.1	& 38.0	& 2.5 \\ 
11264 & 2.5	& 0.4	& 6.3	& 50.5	& 37.9	& 2.3 \\ 
1651	& 3.0	& 0.2	& 6.4	& 56.7	& 31.4	& 2.2 \\ 
1279	& 3.6	& 0.3	& 8.1	& 52.9	& 32.3	& 2.7 \\ 
934	& 4.1	& 0.3	& 8.9	& 47.5	& 36.2	& 3.0 \\ 
644	& 6.2	& 0.5	& 12.1	& 32.6	& 44.4	& 4.2 \\ 
689	& 6.6	& 0.6	& 12.4	& 30.2	& 45.7	& 4.4 \\ 
367	& 12.2	& 0.9	& 13.1	& 16.3	& 52.8	& 4.6 \\ 
179	& 12.0	& 2.8	& 8.3	& 27.8	& 45.6	& 3.6 \\ \hline
\end{tabular}
\caption{The XEDS quantification results for all the elements present corresponding to the V/III series measurement plotted in Fig. 5 (b).}
\label{tableVIII}
\end{table}

At the end of the V/III series experiment, during the XEDS at V/III=179, the nanowire changed direction and folded back onto itself. This gave rise to the relatively high As\% observed at V/III=179 (Ga=52~\%, Au=36~\% and As=12~\% with normalization Au+Ga+As=100\%) Assuming that 12~\% As and hence 12~\% Ga came from scattering from the GaAs nanowire, and renormalizing gives Ga=52.5~\%, Au=47.5~\% which is plotted in Fig. 5 (b).
At the beginning of this particular XEDS scan the catalyst composition measured was Ga=51~\%, Au=45~\% and As=4~\% (with Au+Ga+As=100~\%). In similar experiments conducted there was no additional As at low V/III, and Ga\% always increased at low V/III ratio. An example of another V/III series experiment is shown in the next section.

\section{XEDS at 500~$^\circ$C}

\begin{figure}[!h]
\centerline {\includegraphics[width=170mm]{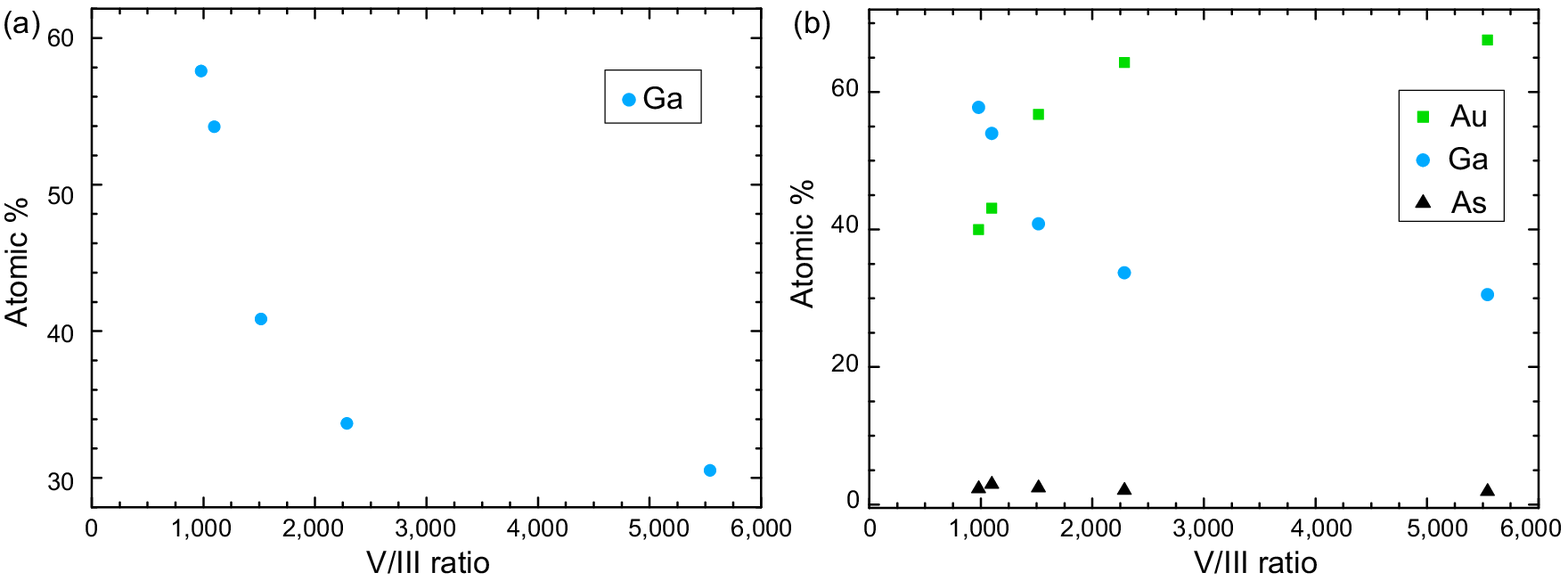}}
\caption{\label{fig:500VIII}(a) The Ga~\% in the catalyst of a nanowire measured at 500~$^\circ$C. (b) The Au and As~\% plotted along with the Ga \% data shown in section (a). In both these plots Au+Ga+As is normalized to 100\%.}
\end{figure}

The catalyst composition at 500~$^\circ$C as a function of the V/III ratio is shown below (Fig. \ref{fig:500VIII}). As expected we see an increase in Ga with decreasing V/III ratio (or increasing TMGa flow.) The As content remains negligible all throughout.
(This experiment was performed in the constant resistance mode of Blaze and the XEDS spectra were acquired for 2 minutes each.)

\section{Phase diagram with As~=~0.3~\%}

\begin{figure}[!h]
\centerline {\includegraphics[width=88mm]{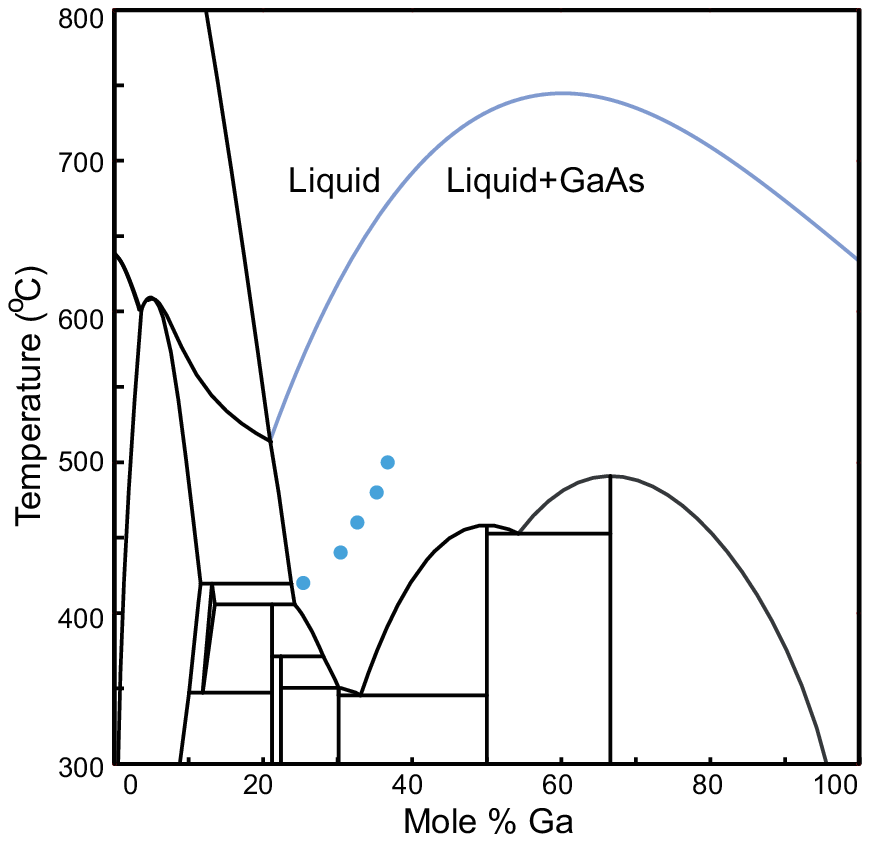}}
\caption{ \label{fig:PDhighAs} Au$_{99.7\%}$As$_{0.3\%}$- Ga$_{99.7\%}$As$_{0.3\%}$  phase diagram (or projection of the Au-Ga-As phase diagram at a fixed As content of 0.3~\%). The blue curve denotes the liquidus. Blue dots are the experimentally measured Ga content as a function of temperature. (Same experimental data as in Fig. 2 (b) of the main article). }
\end{figure}	

Phase diagram section calculated for a fixed As concentration of 0.3~\% overlaid with Ga\% measured during the temperature series (Same experimental data as plotted in Fig. 2 (b) of the article). 

\section{Ga from XEDS and volume change}
As mentioned earlier, the volume of the catalyst particle can give an indirect estimate of the catalyst composition. This is a simple way, particularly useful if the TEM does not have any associated system for quantitative compositional analysis. However, this method assumes that Au is not diffusing out of the catalyst particle, which need not be true at all conditions. In cases where Au diffuses out of the catalyst, this indirect measurement becomes inaccurate. Very evident Au diffusion along Si nanowires during \textit{in situ} experiments have been reported.\cite{hannon2006influence} We have not noticed any obvious migration of Au in our videos, but it is in principle possible at high temperatures.\cite{otnes2016strategies} 
In some experiments (not discussed in this article) we observed tiny portions of the catalyst left on the sidewalls of the nanowire, which we do not know if it was pure Au or pure Ga or a Au-Ga mixture. Since the As concentration is very small, it’s contribution to the volume can be neglected. By correlating the volume change to the catalyst composition measured by XEDS, one can assess Au diffusion on GaAs nanowires at typical growth conditions.

\begin{figure}
\centerline {\includegraphics[width=180mm]{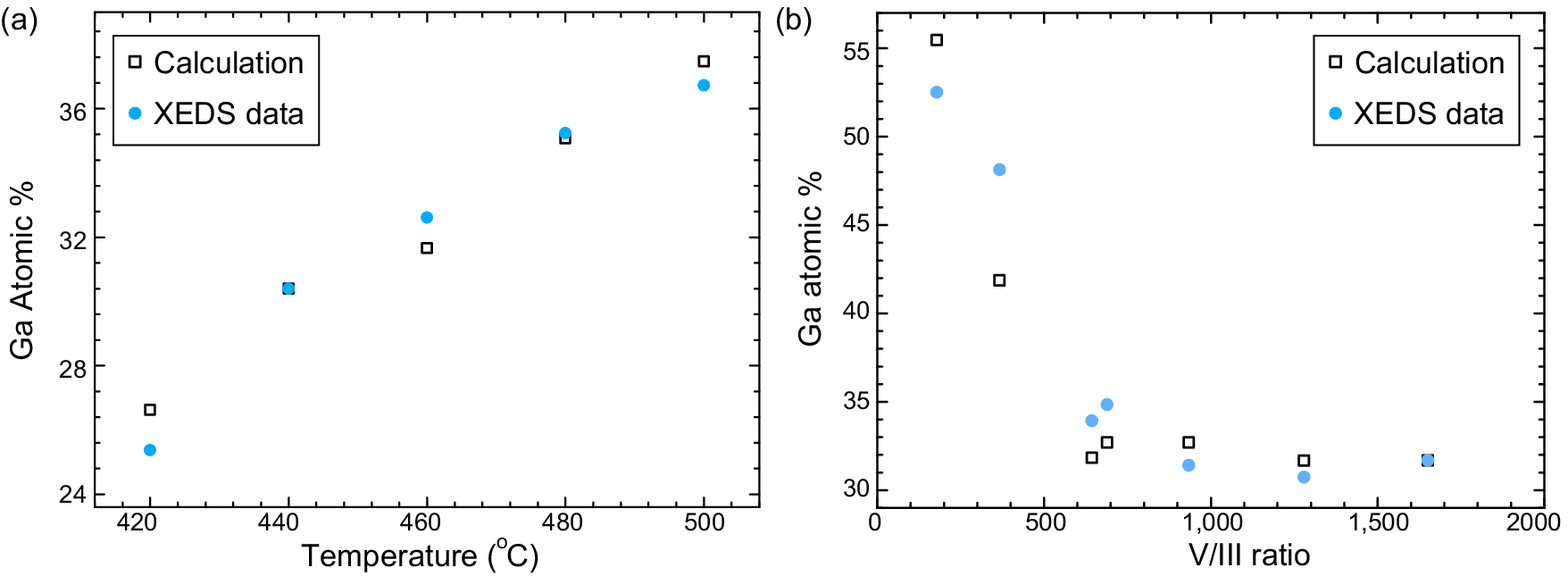}}
\caption{\label{fig:voltoGa}Estimating catalyst composition from catalyst volume change: The Ga\% calculated from the measured dimensions of the catalyst particle is shown by the open black squares. The Ga measured by XEDS is shown by the filled blue circles. (a) is for the temperature series discussed in the article as Fig. 2 (b). The normalization is done with XEDS and volume measured at 440~$^\circ$C. (b) is for the V/III series measured at 420~$^\circ$C and discussed in the main article as Fig. 5 (b). The normalization is performed at V/III=1651.}
\end{figure}

	The Ga concentration in the catalyst calculated from the volume for the temperature series experiment discussed earlier is plotted in Fig. \ref{fig:voltoGa} (a) as open black squares. For this calculation we have imposed that the volume measured at 440~$^\circ$C is due to 30~\% Ga, which is the XEDS value measured at 440~$^\circ$C. The volume was calculated by treating the catalyst particle as a spherical cap and assuming the base of the catalyst to be a circle. The measured and calculated Ga\% matches, indicating that the Au diffusion is negligible.  (From the catalyst volume and Ga content measured during growth, the diameter of the starting Au seed particle can be calculated to be about 28~nm, assuming the starting Au seed would have been spherical. The exact dimensions of the catalyst giving this specific nanowire that was measured is unknown to us. But the average size of the Au particles deposited on the substrate during this experiment was about 30 nm in diameter, which agrees to the 28 nm we calculated for this particular nanowire.)
Similarly, Ga\% calculated from the volume for the V/III series is plotted in Fig. 5 (b). The normalization was done by setting the volume measured at V/III = 1651 to be due to the measured value of 31~\% Ga. The catalyst shape was fitted with a prolate spheroid with the nanowire-catalyst interface being a circle.

\end{document}